# High-performance orbital-torque magnetic memory on the 300-mm platform

Dinggui Zeng[1,†,*], Yang Gao[1,†], Jinyu Duan[2,†], Lei Zhao[1], Yuhao An[2], Xing He[3], Jintao Ke[3], Yonglong Ga[4], Shasha Wang[1], Zhenghui Ji[1], Muyuan Chen[1], Hengan Zhou[1], Xuejie Xie[1], Enlong Liu[1], Junlu Gong[1], Qijun Guo[1], Yihui Sun[1], Zejie Zheng[1], Weiming He[1], Xiaolei Yang[1], Fantao Meng[1], Yaohua Wang[1], Hongxin Yang[4], Delin Zhang[2,*], Yong Jiang[2], Wanjun Jiang[3,*], Shikun He[1,*]

1. Zhejiang Hikstor Technology Co. LTD., 311305, Zhejiang, China.
2. Institute of Quantum Materials and Devices; School of Electronic and Information Engineering, Tiangong University, Tianjin, China.
3. State Key Laboratory of Low-Dimensional Quantum Physics and Department of Physics, Tsinghua University, 10084, Beijing, China.
4. Center for Quantum Matter, School of Physics, Zhejiang University, 310058, Zhejiang, China.

†These authors contributed equally: Dinggui Zeng, Yang Gao, and Jinyu Duan

*Corresponding authors. Email: zengdinggui@hikstor.com (D.G.Z.); zhangdelin@tiangong.edu.cn (D.L.Z.); jiang_lab@mail.tsinghua.edu.cn (W.J.J); he_shikun@hikstor.com (S.K.H.)

**Abstract: Contemporary memory technologies are increasingly constrained by the fundamental trilemma of storage capacity, access latency, and power consumption. Among the emerging technologies, spin-orbit torque magnetic random-access memory (SOT-MRAM) shows promise to circumvent these challenges, owing to its fast switching dynamics and high endurance. However, the application of SOT-MRAM is hindered by the relatively low write and read efficiencies, resulting in a large bitcell area and an insufficient sensing margin. Meanwhile, the involvement of an ultrathin spin-source channel, typically within a few nanometers, imposes technological challenges for mass production. Here, we resolve these issues on a 300-mm wafer platform by exploiting the emerging orbital degree of freedom and the resultant orbital torque (OT) from the relatively thick Ti/W bilayer. In particular, OT memory nanodevices exhibit a giant tunnel magnetoresistance (TMR) of 182%, nanosecond-scale response, $10^{12}$ endurance, together with an enhanced switching efficiency ($E_b/I_c$), which consequently enables an ultra-low write energy of less than 0.1 pJ/bit. Our findings demonstrate that orbital angular momentum can be implemented for building energy-efficient MRAM devices, offering a practical pathway towards low-latency memory that is demanded for high-performance computing and AI applications.**

Modern computing increasingly demands high-performance memory technology that combines low latency, high density, high endurance and low energy consumption[1,2]. Charge-based memory devices, such as static random-access memory (SRAM), provide fast access and high throughput and are therefore widely used in artificial-intelligence (AI) accelerators[3], but their volatility and large bitcell area limit further scaling[4]. Emerging non-volatile memories seek to overcome these constraints[5-8]. Among them, spin-based memory, spin-orbit torque magnetic random access memory (SOT-MRAM) in particular, offers fast switching speed and high endurance[9-13], but its density and read performance remain limited by two correlated factors. Despite extensive efforts[14-19], the state-of-the-art SOT-MRAM devices typically require a large write current ($\geq$ 400 μA)[20], which increases the bitcell area by increasing the size of the access transistor in a two-transistor-one-magnetic-tunnel-junction (2T-1MTJ) architecture[21]. At the same time, the reported tunnel magnetoresistance (TMR) ratios are in the range of 80% to 156%[22-25], restricting the sense margin and impeding high-speed readout in last-level cache applications[26,27].

Recent studies have identified orbital angular momentum as an alternative route for generating current-induced torque and for promoting MRAM technology[28-30]. In low-cost light metals, a flow of charge current can produce transverse orbital currents, which can be converted into spin current and exert an orbital torque (OT) on the adjacent magnets[31-39]. Such an OT could reduce the write current in MRAM devices. Meanwhile, the relatively thicker orbital-current channels[40-45] offer a thickness tolerance than conventional ultrathin spin-source layers, providing a potentially larger process window for scalable production. Towards OT-MRAM devices[46], a simultaneous realization of a low write current, a large TMR ratio, a nanosecond operation, and compatibility with advanced manufacturing processes remain to be demonstrated. Here we address these challenges by optimizing the orbital-current channel of stacking order Ti/W, together with the integration of CoFeB-based perpendicular MTJ (pMTJ) nanodevices on a 300-mm wafer platform. In particular, we realize deterministic perpendicular magnetization switching with a large OT efficiency of $\xi_{OT}$ ~ 0.33 and a low critical switching current density of $J_c$ ~ $5 \times 10^6$ A/cm$^2$ in Hall bar devices. In an array of three-terminal pMTJ nanodevices on 300-mm platform, we demonstrate a fast operation speed at 2 ns, together with a reduced critical switching current down to ~ 250 ± 30 μA, corresponding to an ultra-low write energy below 0.1 pJ/bit. More importantly, a large TMR ratio up to 182% is obtained, supplemented by a sufficient read sense margin in a wide temperature range (25 °C – 125 °C). These results suggest that Ti/W based orbital-current channel can jointly improve write efficiency, read performance and process tolerance in nanoscale magnetic memory, offering a compelling alternative for low-latency and energy-efficient emerging magnetic memories.

## Giant TMR and large OT efficiency

**Figure 1a** illustrates the optimized pMTJ stack, in which a Ti/W bilayer serves as the orbital-current channel, which is subsequently integrated with the CoFeB-based pMTJ stack (**Details in Methods section**). The complete OT-MRAM stack is deposited on a 300-mm wafer. For comparison, otherwise equivalent pMTJ stacks and devices, incorporating W, Ru/W and Mo/W are made following the same process conditions. To survey the TMR ratios, we first employ the current-in-plane tunnel (CIPT) method. The Ti/W-based pMTJ stack exhibits a maximum TMR ratio of 205% at a resistance–area product (RA) of about 95 Ω·μm$^2$, exceeding the values obtained from the W-based, Ru/W-based and Mo/W-based control stacks by approximately 40-50% (**Fig.**

**1b and Supplementary Note 1**). These results identify the unique role of Ti/W bilayer for enabling a high-quality MTJ.

To elucidate the structural origin of the enhanced TMR ratio, we examine the Ti/W-based and W-based pMTJ stacks using scanning transmission electron microscopy (STEM). **Figures 1c** and **1d** present the cross-sectional high-angle annular dark-field (HAADF)-STEM images. In the Ti/W-based stack, the Ti layer exhibits a textured hexagonal-close-packed (*hcp*) structure with a (0001) out-of-plane orientation, whereas the subsequently deposited W layer adopts the $\alpha$-phase with a (110) orientation, as shown by HAADF-STEM imaging and corresponding fast Fourier transform (FFT) analysis (**Fig. 1c**). The (0001) orientation of the Ti layer is further supported by a pronounced (002) diffraction peak from the X-ray diffraction (XRD) measurement (**Extended Data Fig. 1a**). Based on the stacking order of Ti (0001)||W (110), a lattice mismatch approximately 7% can be estimated (**Extended Data Fig. 1b**).

The textured growth of the Ti/W bilayer promotes the crystalline ordering in the overlying CoFeB/MgO layers, providing a structural basis for the enhanced TMR ratio[47]. Chemically well-defined interfaces throughout the multilayer stack can be seen from the elemental mapping, which is obtained via energy-dispersive X-ray spectroscopy (EDS) (**Extended Data Fig. 1c**). Through electron-energy-loss-spectroscopy (EELS) analysis, it is found that the Ti layer serves as a boron sink due to its relatively large negative formation enthalpies of titanium borides[48] (**Extended Data Fig. 1d and Data Table 1**). Boron removal from CoFeB facilitates the crystallization of the CoFe and MgO barrier, thereby improving the symmetry-selective tunnel process responsible for large TMR[47]. By contrast, the W-based stack deposited directly on $SiO_2$ wafer exhibits a polycrystalline structure, and less pronounced structural ordering in the CoFeB/MgO region (**Fig. 1d**), resulting in a reduced TMR.

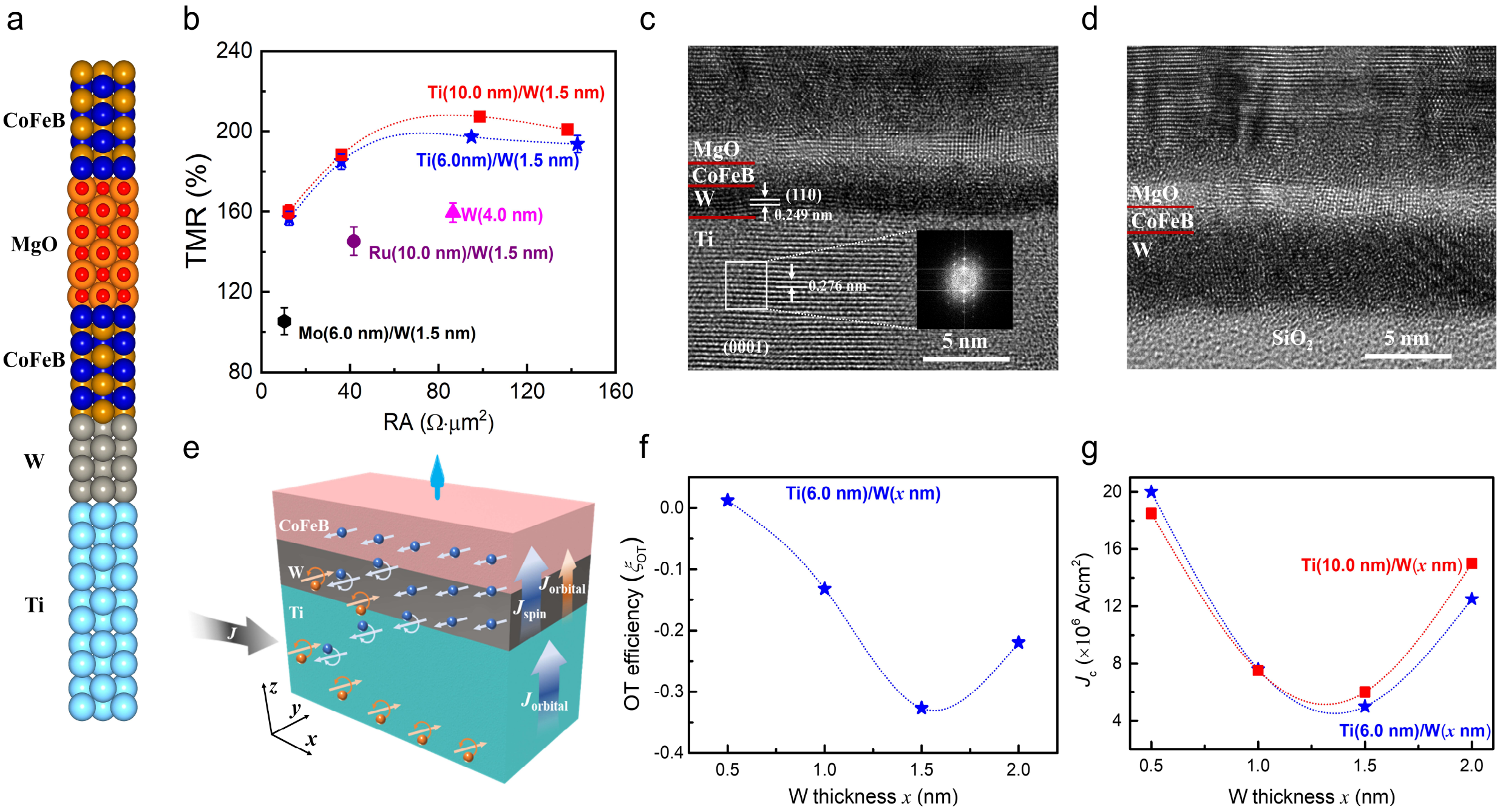

**Fig. 1 | Multilayer stacks with optimized TMR and OT efficiency. a** Schematic of the Ti/W-based pMTJ stack that accommodates the large OT and enhanced TMR. **b** The TMR ratio as a function of the resistance-area product (RA) for the Ti/W, W, Ru/W, and Mo/W-based stacks that are measured with current-in-plane-tunnel (CIPT) method. Error bars represent the standard deviation of the measured TMR over mean. **c,d** The cross-sectional high-angle annular dark field scanning transmission electron microscope (HAADF-STEM) images of the Ti/W-based pMTJ and W-based pMTJ stacks, respectively. Inset to **c** shows the Fast-Fourier-Transform (FFT) image from the selected region of the Ti layer. **e** Schematic of orbital-current generation in the Ti/W bilayer and its conversion into a torque acting on the CoFeB layer. **f** OT efficiency ($\xi_{OT}$) as a function of W thickness in Ti (6 nm)/W ($t_W$ = 0.5, 1.0, 1.5, and 2.0 nm)/CoFeB (5.0 nm)/MgO (1.0 nm)/Ta (2.0 nm) multilayers, which is determined by using the ST-FMR method. **g** Critical

switching current density ($J_c$) as a function of $t_W$ for Ti ($t_{Ti}$ = 6, 10 nm)/W ($t_W$ = 0.5, 1.0, 1.5, and 2.0 nm)/CoFeB (1.05 nm)/MgO (1.0 nm)/Ta (2.0 nm) Hall bar devices.

The OT efficiency ($\xi_{OT}$) is theoretically examined in the Ti/W-based multilayer. First-principles calculations indicate that *hcp* Ti exhibits a large orbital Hall conductivity [$\sigma_{OHE}$ ~ 3728 (ħ/e)(S/cm)], but a much smaller spin Hall conductivity [$\sigma_{SHE}$ ~ 25 (ħ/e)(S/cm)] (**Extended Data Fig. 2c**). These results support that the charge current in Ti layer generates predominantly orbital current. The orbital angular momentum is then transmitted into the adjacent $\alpha$-phase W layer, where the spin-orbit coupling enables the orbital-to-spin conversion. Note that the $\alpha$-phase W exhibits pronounced orbital Hall conductivities of $\sigma_{OHE}$ ~ 4885 (ħ/e)(S/cm) (**Extended Data Fig. 2d**) and spin Hall conductivity $\sigma_{SHE}$ ~ 748 (ħ/e)(S/cm), which could further promote the current-induced magnetization switching[36,49], as schematically shown in **Fig.1e**. This charge-to-orbital-to-spin conversion process provides the microscopic basis for the enhanced current-induced torque illustrated in **Fig. 1e**.

The OT efficiency ($\xi_{OT}$) is experimentally quantified by implementing the spin-torque ferromagnetic resonance (ST-FMR) technique[34] (**Details in Methods section** and **Extended Data Fig. 3**). **Figure 1f** presents the estimated $\xi_{OT}$ values for Ti (6 nm)/W ( $t_W$ nm)/CoFeB (5 nm)/MgO/Ta (with $t_W$ = 0.5, 1.0, 1.5, and 2.0 nm). The magnitude of $\xi_{OT}$ initially increases with W thickness, reaches a maximum around 0.33 at $t_W$ = 1.5 nm, and decreases to 0.23 upon further increasing the thickness of the W layer at $t_W$ = 2 nm. The current-induced perpendicular magnetization switching is presented in both Ti (6 nm)/W ($t_W$ = 0.5, 1.0, 1.5, and 2.0 nm)/CoFeB (1.05 nm)/MgO/Ta and Ti ($t_{Ti} =$ 4, 6, 8, 10 nm)/W (1.5 nm)/CoFeB (1.05 nm)/MgO/Ta Hall bar devices. The room-temperature anomalous Hall resistances ($R_{xy}$) as a function of out-of-plane field ($H_z$) exhibit a square shape, manifesting the presence of perpendicular magnetic anisotropy (PMA) (**Extended Data Figs. 4a-l**). Under in-plane fields along the $x$ axis $H_x = \pm 10$ mT, the sweeping of current pulse ($I_{pulse}$) could deterministically switch the low-$R_{xy}$ and high-$R_{xy}$ states. Reversing the sign of in-plane fields ($\pm H_x$) reverses the polarity of the $R_{xy}$ $vs.$ $I_{pulse}$ loops, consistent with the current-induced 180° perpendicular magnetization switching[42-45].

The role of the W layer in governing the critical switching current density ($J_c$) is investigated. For a fixed Ti thickness of $t_{Ti}$ = 6 nm, $J_c$ decreases following the increase of $t_W$, and reaches a minimum value of $J_c = 5.0 \times 10^6$ A/cm$^2$ at $t_W$ = 1.5 nm (**Fig. 1g**). With $t_W$ = 1.5 nm being fixed, the varied Ti thicknesses ($t_{Ti}$) from 4 nm to 10 nm result in a relatively weak change of $J_c$ (4.5 – 6.0 × 10$^6$ A/cm$^2$) (**Extended Data Fig. 4m**). Note that the W-thickness dependence of $J_c$ closely follows the trend of the effective OT efficiency ($\xi_{OT}$) shown in **Fig.1f**, with both quantities being optimized near $t_W$ = 1.5 nm. This correlation indicates that the W thickness plays a central role in determining the efficiency of the orbital-to-spin conversion ($\xi_{OT}$) and consequently the critical current density ($J_c$) required for perpendicular magnetization switching.

**Read performance of OT pMTJ nanodevices**

Following the optimization of TMR ratio and OT responses, a Ti (6.0 nm)/W (1.5 nm)-based pMTJ stack, together with a W (4.0 nm)-based control stack are integrated into the standard 300-mm back-end-of-line (BEOL) platform (**Fig. 2a**). Arrays of three-terminal pMTJ nanodevices are patterned with a lateral dimension of 60 nm (width) × 130 nm (length) by employing the self-aligned rounded rectangular (SARR) integration technique[50]. This integration scheme reduces alignment complexity and supports a high device yield (**Supplementary Note 2 and Supplementary Fig. 2**). **Figure 2b** presents a cross-sectional transmission electron microscopy (TEM) image of the Ti/W-based pMTJ nanodevice, where the MTJ, bottom via (BV), top electrode (TE) and top via (TV) interconnects are clearly resolved.

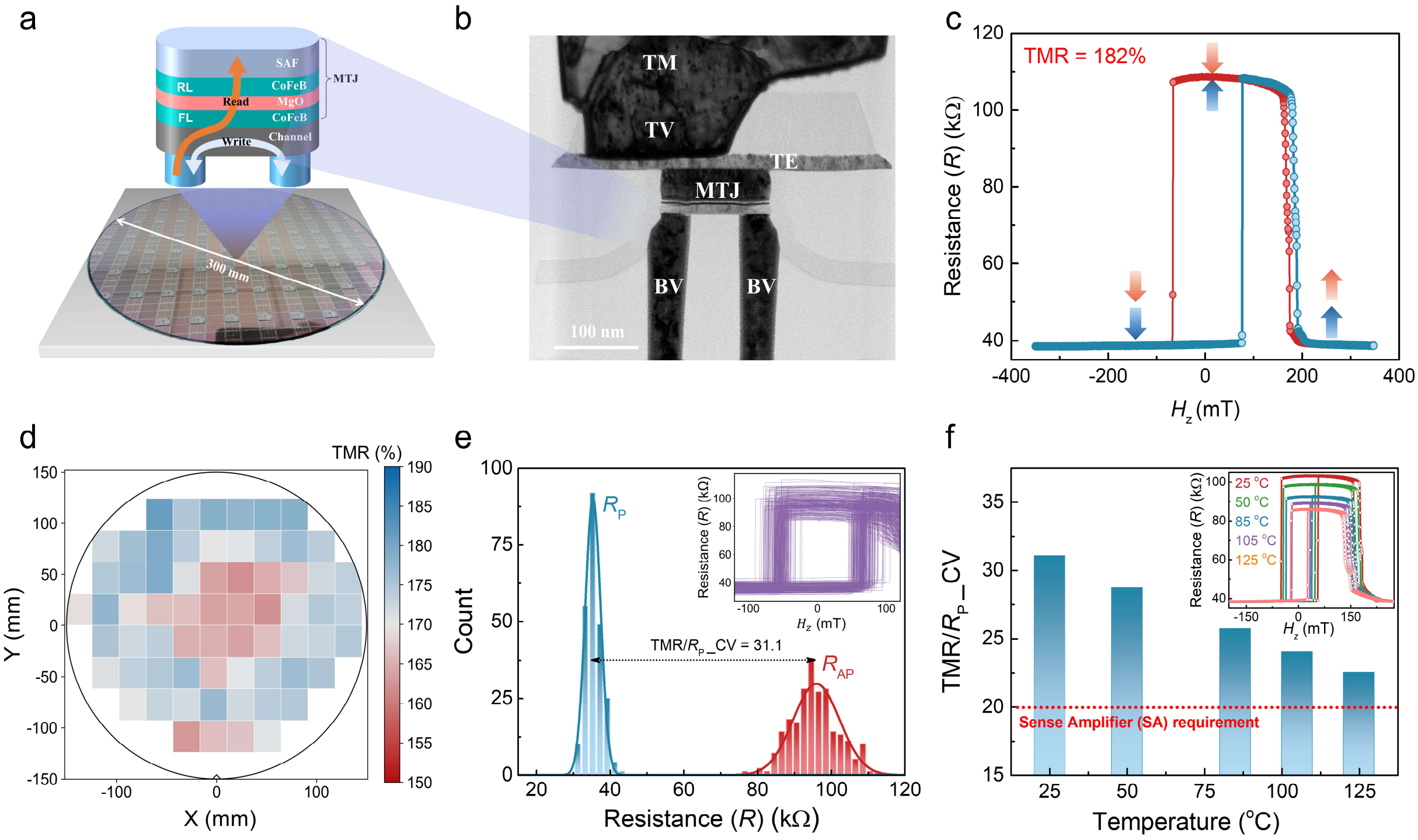

**Fig. 2 | Wafer-scale read performance of Ti/W pMTJ nanodevices. a,b** Schematic and cross-sectional TEM image of the Ti/W-based OT pMTJ nanodevices with lateral dimensions of ~60 nm × 130 nm, which are fabricated on 300 mm wafers with the bottom via (BV), top via (TV), top metal (TM) and top electrode (TE). Note that RL denotes the reference layer, while FL denotes the free layer. Directions of arrows in (**a**) show the electrically separated read and write paths. **c** Major loop of the resistance versus perpendicular magnetic field ($R-H_z$) of a pMTJ nanodevice,

showing a TMR ratio up to 182%. **d** Wafer-scale map of the TMR ratio. **e** Distributions of parallel-state ($R_P$) and antiparallel-state ($R_{AP}$) resistances, respectively. The inset shows minor $R{-}H_z$ loops (in a field range of $\pm 100$ mT) measured from 200 randomly selected pMTJ nanodevices. **f** Temperature dependence of the wafer-averaged read margin metric, TMR/$R_P$_CV for Ti/W-based OT pMTJ nanodevices (from 25 °C to 125 °C). Here, $R_P$_CV denotes the $R_P$ coefficient of variation, statistically derived from the standard deviation of $R_P$ over mean. Inset image shows the measured major $R$-$H_z$ loops for temperatures between 25 °C to 125 °C.

The major loop ($R$-$H_z$) of a representative Ti (6.0 nm)/W (1.5 nm)-based pMTJ nanodevice is shown in **Fig. 2c**. The reference layer (RL) and free layer (FL) switch at approximately 160 mT and 60 mT, respectively, providing a clear separation window between them. The high- and low-resistance states that correspond to the antiparallel ($R_{AP}$) and parallel ($R_P$) configurations are clearly observed, yielding a TMR ratio $[(R_{AP} - R_P)/R_P \times 100\%]$[51] of up to 182%. By measuring over 200 pMTJ nanodevices, the averaged TMR ratio can be estimated as 171% $\pm$ 7% (**Fig. 2d**). By contrast, a smaller TMR ~ 130% is observed in W-based pMTJ nanodevices, as shown in **Extended Data Fig.5.**

Wafer-scale device uniformity is reflected by a die-to-die variation of 2.9% of the averaged TMR ratio (**Fig. 2d**). More importantly, the read margin, quantified by the figure of merit TMR/$R_\mathrm{P}$_CV (where $R_\mathrm{P}$_CV is the coefficient of variation of $R_P$), surpassing 31 at room temperature (**Fig. 2e**), with $R_\mathrm{P}$_CV ~ 5.5% in Ti (6.0 nm)/W (1.5 nm)-based pMTJ nanodevices. For high temperature applications, the pMTJ nanodevices need to maintain a stable read margin[26] of TMR/$R_\mathrm{P}$_CV > 20. The temperature dependence of TMR ratio (from 25 °C to 125 °C) is shown in **Extended Data Fig. 5c**. A large TMR/$R_\mathrm{P}$_CV value up to ~ 26 at 85 °C and ~ 23 at 125 °C can be seen, thereby satisfying the sense-amplifier requirement for ensuring reliable and high-speed read operations[26], as shown in **Fig. 2f**.

**Write characteristics of OT pMTJ nanodevices**

**Figure 3a** shows the resistance versus the write voltage (*R*-*V*) of Ti (6.0 nm)/W (1.5 nm)-based pMTJ nanodevices, measured using pulse widths $t_{pulse}$ = 2 ns, 5 ns, and 10 ns, respectively. Clear antiparallel (AP)-to-parallel (P) and P-to-AP switching behaviors are observed upon reversing the polarity of write voltage (*V*) that is applied along the Ti/W channel. Note that the *R*–*V* loops show a small voltage offset, as a result of the stray fields of the top SAF layer. At a pulse width of 2 ns and under the in-plane field $H_x$= -35 mT, a deterministic switching is achieved under critical voltage $V_c$ ~ 0.18 V. Based on $I_c = V_c/R_{ch}$ (where $R_{ch}$ is the Ti/W channel resistance), we obtain a critical write current ($I_c$) of approximately 232 μA in the pMTJ nanodevice. The corresponding write energy $E_{sw} = I_c^2 \times R_{ch} \times t_{pulse}$, can be computed as 0.08 pJ/bit. To evaluate the switching probability ($P_{sw}$), we perform 100 consecutive *R*-*V* sweeps at $t_{pulse}$ = 2 ns, 5 ns and 10 ns, respectively. As shown in **Fig. 3b**, $P_{sw}$ reaches 100% for both AP-to-P and P-to-AP switching at a low write voltage below 0.23 V. These voltages are below the reported values of ~ 0.4 V[20,52] for representative SOT-MRAM devices and CMOS power supply voltage (e.g., ~ 0.7 V at advanced 7 nm tech node)[53], demonstrating the low-voltage, low-power, nanosecond switching characteristics and the potential scalability of the Ti/W-based pMTJs.

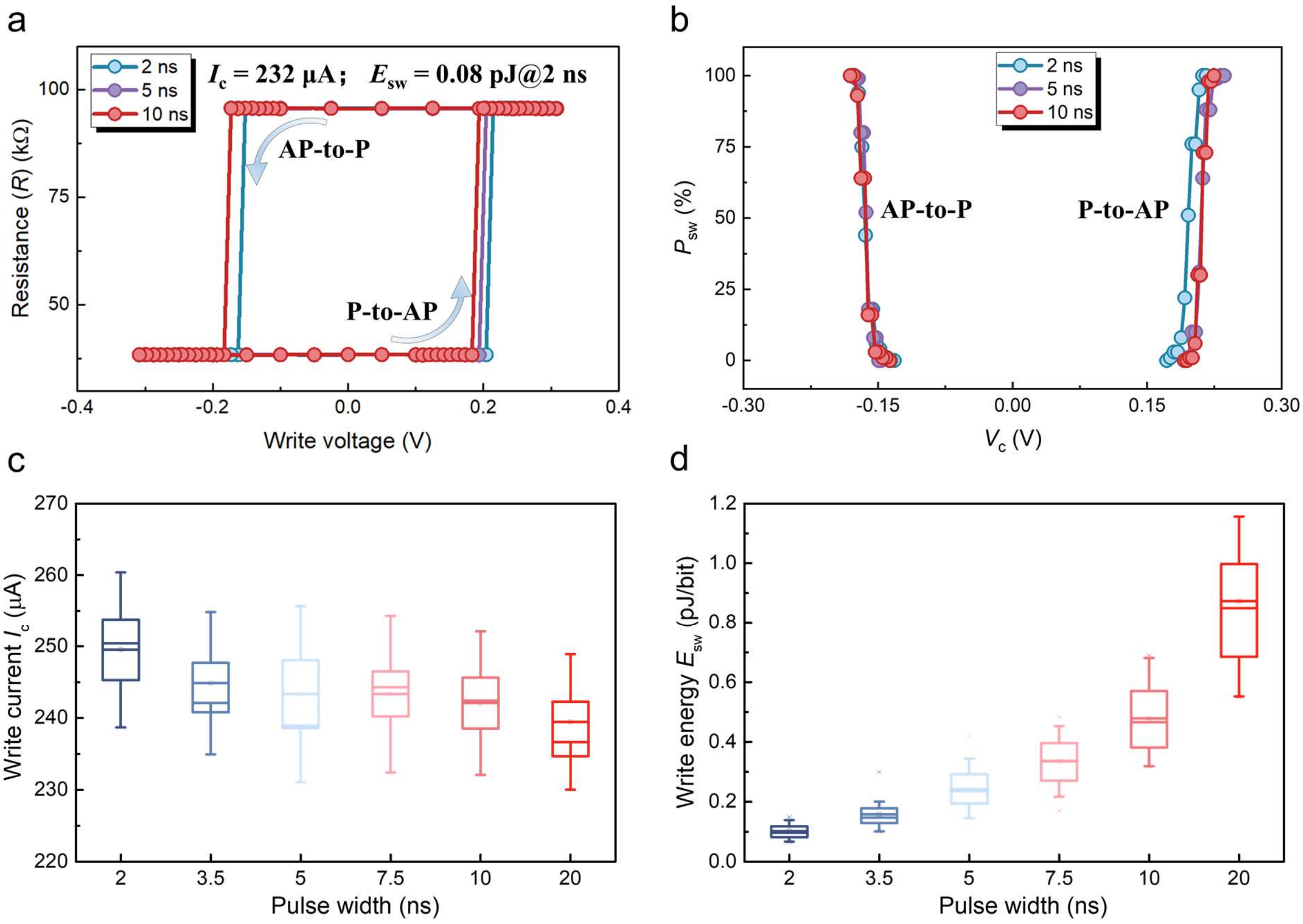

**Fig. 3 | Write characteristics of OT pMTJ nanodevices. a** The pMTJ resistance versus write voltage ($R$-$V$) loops of Ti (6.0 nm)/W (1.5 nm)-based pMTJ nanodevices under $t_{pulse}$ = 2 ns, 5 ns, and 10 ns. An in-plane magnetic field $H_x$= -35 mT is applied along the voltage/current direction in the Ti/W channel. **b** Switching probability ($P_{sw}$) as a function of critical write voltage ($V_c$) for $t_{pulse}$ = 2 ns, 5 ns, and 10 ns evaluated by repeating 100 switching experiments. **c,d** Critical switching current ($I_c$) and estimated write energy ($E_{sw}$) of the Ti/W channel as a function of pulse width $t_{pulse}$. These values are averaged over antiparallel-to-parallel (AP-to-P) and parallel-to-antiparallel (P-to-AP) resistance states.

The evolution of $I_c$ and $E_{sw}$ as a function of pulse width $t_{pulse}$ is plotted in **Figs. 3c and 3d**, respectively. Following the increase of $t_{pulse}$, $I_c$ increase slightly, whereas $E_{sw}$ decreases monotonically. Such an observation can be understood from the reduced values of $t_{pulse}$ that overweight the modest increase of $I_c$ in the 2 to 20 ns range[30], which can be inferred from $E_{sw} = I_c^2 \times R_{ch} \times t_{pulse}$. At $t_{pulse}$= 2 ns, the write energy $E_{sw}$ of Ti (6.0 nm)/W (1.5 nm)-based pMTJ nanodevices could reach as low as ~ 0.1 pJ/bit. In comparison with W (4.0 nm)-based control pMTJs, the Ti/W-based pMTJ nanodevices show a substantial reduction of $I_c$ (~ 34%) and $E_{sw}$ (~ 44%) at $t_{pulse}$= 2 ns (**Extended Data Figs. 6a-6b**).

The write energy is inherently tied to the thermal stability. As such, the coercivity ($H_c$) of the FL of pMTJ and the thermal stability factor ($\Delta$) of pMTJ nanodevices are studied[25]. The thermal stability factor is defined as $\Delta = E_b/k_B T$, where $E_b$ is the energy barrier and $k_B$ is the Boltzmann constant. The coercivity ($H_c$) ~ 63 mT is observed in the Ti (6.0 nm)/W (1.5 nm)-based pMTJ nanodevices, which is larger than that of ~ 48 mT for the W (4.0 nm)-based pMTJ nanodevices (**Extended Data Fig. 6c**). This increase is consistent with enhanced interfacial quality in the Ti/W bilayer, which enhances the interfacial PMA at the CoFe/MgO interface[54]. For Ti/W-based pMTJ nanodevices, a thermal stability factor $\Delta$ ~ 58 is extracted, which is extrapolated using the magnetic field acceleration method by employing domain wall fitting[55,56] (**Details in Methods section**). In contrast, a smaller $\Delta$ ~ 47 is observed in the W-based control nanodevices (**Extended Data Fig. 6d and Data Fig. 7**). As a result, a room-temperature retention time exceeding 10 years can be projected in Ti/W-based pMTJ nanodevices.

## Performance metrics of OT-MRAM nanodevices

We collectively evaluate the read stability, write efficiency, and operational robustness of the Ti (6.0 nm)/W (1.5 nm)-based OT-MRAM pMTJ nanodevices. The key device metrics are summarized in **Fig. 4 and Table 1**. As compared with the W (4.0 nm)-based nanodevices, the switching efficiency, defined as $E_b/I_c$, is improved from ~ 0.13 $k_B T/\mu A$ to ~ 0.23 $k_B T/\mu A$ (**Fig. 4a**), exhibiting a 77% enhancement, highlighting the improved write efficiency by using Ti/W OT channel. Furthermore, endurance is evaluated by periodically applying bipolar write pulses up to $10^{12}$ cycles. The parallel and antiparallel resistance states ($R_P$ and $R_{AP}$), together with the channel resistance ($R_{ch}$), show no detectable degradation, as shown in **Fig. 4b**. Concurrently, statistics over more than 2,000 devices across the whole wafer reveal a device yield exceeding 99.95% (**Extended Data Fig. 8**), suggesting a large process tolerance that are beneficial for mass and scalable production. Benchmarking against representative SOT[20,25,57-61] and OT-MRAM[30] devices reported in the literature, the Ti (6.0 nm)/W (1.5 nm)-based OT-MRAM nanodevices exhibit improved read and write performance metrics. These parameters, including TMR ratio, TMR/$R_P$_CV, $I_c$ as well as $E_{sw}$, are tabulated in **Table 1**.

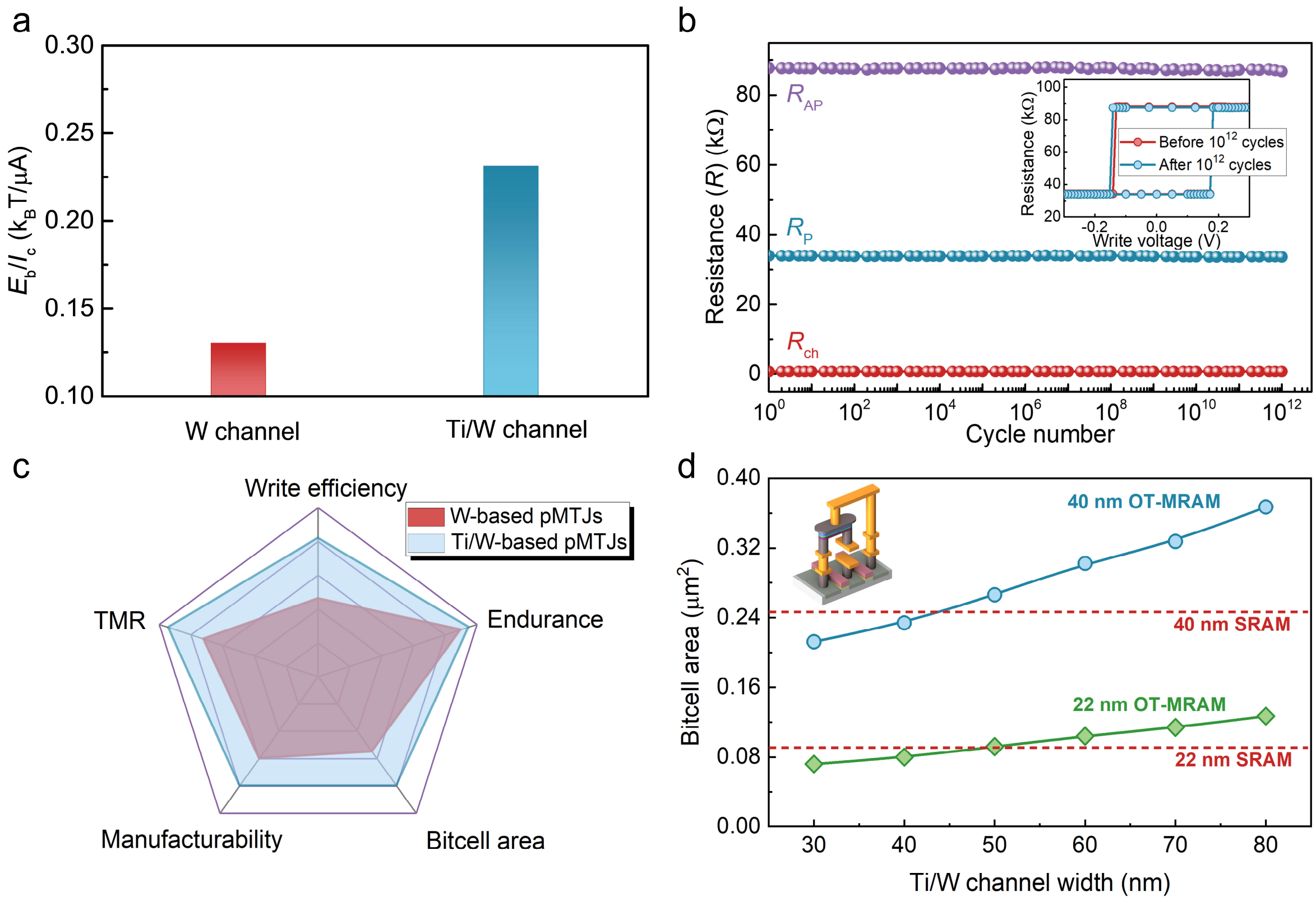

**Fig. 4 | Device-level performance metrics and projected scalability of the OT-MRAM nanodevices. a** Comparison of switching efficiency ($E_b/I_c$) of Ti (6.0 nm)/W (1.5 nm)-based OT-MRAM pMTJs against W (4.0 nm)-based SOT-MRAM pMTJs nanodevices. $E_b/I_c$ is a figure of merit for write efficiency. **b** Parallel, antiparallel and channel resistances ($R$) as a function of cycle number ($N$) of the Ti (6.0 nm)/W (1.5 nm)-based pMTJ nanodevices, tested with periodic bipolar voltage pulses of 5 ns and 0.3 V (applied approximately 1.5 times larger than $V_c$ to cover wafer-level device-to-device variations). The Inset shows representative pMTJ resistance versus write voltage ($R$-$V$) loops measured before and after $10^{12}$ switching cycles. **c** Performance metrics of the Ti/W against W-based pMTJ nanodevices, including TMR, write efficiency, endurance, process tolerance and projected bitcell area. **d** Simulated bitcell area of the OT-MRAM 2T-1MTJ (two-transistor and one-magnetic-tunnel-junction) architecture as a function of OT-channel width,

varied from 30 nm to 80 nm, for 40 nm and 22 nm technology nodes. The inset depicts the corresponding 2T-1MTJ bitcell layout.

| Parameters | | This work | Hikstor (Ref. 25) | Beihang (Ref. 30) | TSMC (Ref. 61) | Tohoku (Ref. 60) | Beihang (Ref. 59) | TSMC (Ref. 58) | IMEC (Ref. 20) | Intel (Ref. 57) |
|---|---|---|---|---|---|---|---|---|---|---|
| MTJ type | | PMA | PMA | PMA | In-plane | In-plane | In-plane | In-plane | PMA | PMA |
| Switching origin | | Orbital current | Spin current | Orbital current | Spin current | Spin current | Spin current | Spin current | Spin current | Spin current |
| Wafer scale | | 300 mm | 300 mm | 200 mm | 300 mm | 300 mm | 200 mm | 300 mm | 300 mm | 300 mm |
| Film TMR | | 200% | 198% | N.A. | N.A. | N.A. | N.A. | N.A. | N.A. | 160% |
| MTJ size (nm) | | 60×130 | 80 | 250 | 75×230 | 88×315 | 260×720 | 75×230 | 63 | 57 |
| Device TMR (%) | | 182 (Max); 171 (Ave) | 168 (Max); 156 (Ave) | 89 (Max) | 146 (Max) | 170 (Max) | 80 (Ave) | 140 (Ave) | 86 (Ave) | 127 (Ave) |
| TMR /$R_P$_CV | | ~31 @RT; ~26 @85 ºC | ~29 @RT; ~24 @85 ºC | N.A. | N.A. | N.A. | 25.6 @RT | N.A. | N.A. | N.A. |
| $H_c$ (mT) | | 63.1 | 61.8 | ~35 | ~15 | N.A. | N.A. | ~20 | 90 | ~60 |
| $I_c$ (μA) | 2ns | 249 | 764 | N.A. | N.A. | N.A. | N.A. | N.A. | N.A. | N.A. |
| | 1ns | N.A. | N.A. | N.A. | ~728 | 618 | N.A. | N.A. | 550 | N.A. |
| $E_{sw}$ (pJ/bit) | | 0.10 @2 ns | 0.89 @2 ns | 0.24 @1 ns (projected) | N.A. | 0.16 @1 ns | N.A. | N.A. | 0.21 @1 ns | N.A. |
| Retention Δ | | 57.6 | 56.3 | N.A. | 116 | 70 | 72 | 152 | 53 | N.A. |
| Endurance | | >1E12* | >1E12 | N.A. | N.A. | N.A. | >1E10 | >7E12 | >1E13 | N.A. |
| Device yield (%) | | 99.95% | 99.6% | N.A. | N.A. | N.A. | 99.9% | 98.2% | N.A. | 96% |

(*Actual measurement, limited by endurance test time)

**Table 1 | Benchmark of the performance for the OT- and SOT-MRAM.** Both in-plane and scalable perpendicular MTJ types are included. As compared to previous literatures[20,25,30,57-61], the newly developed Ti/W-based OT-MRAM pMTJ nanodevices exhibit outstanding TMR, read margin, write current, write power, reliability as well as device yield.

**Fig. 4c** summarizes the performance metrics. The Ti (6.0 nm)/W (1.5 nm)-based OT pMTJ nanodevices exhibit clear advantage over the W-based control nanodevices, including the enhanced TMR ratio, write efficiency, and a lower projected bitcell area. Meanwhile, the weak dependence of the critical switching current density on Ti thickness (between 6 and 10 nm) indicates a broader process window than that of conventional ultrathin spin-source layers of the SOT-MRAM (**Supplementary Note 3**). In addition, the smaller write current reduces the drive-current requirement of the access transistors and therefore enables a smaller 2T-1MTJ bitcell area that is beneficial for scaling of MRAM. As shown in **Fig. 4d**, the projected bitcell area of the OT-MRAM with 40 nm-channel width could reach 0.23 $\mu m^2$ at 40-nm node, which is smaller than that of an SRAM cell at the equivalent node[62]. Upon scaling down to 22-nm node, the bitcell area can be further reduced to 0.08 $\mu m^2$. These results indicate that the Ti/W-based OT architecture could support high-density integration at advanced technology nodes. Note that the circuit-level simulations are done by implementing the experimentally measured device parameters in conjunction with the process design kit (PDK) models for 40-nm and 22-nm CMOS technology nodes (**Supplementary Note 4**).

**Discussion and outlook**

In conclusion, we have systematically studied the orbital-torque (OT) efficiency in the Ti/W bilayers, and its integration with the perpendicular magnetic tunnel junctions (pMTJ) on a 300-mm MRAM platform. Specifically, we demonstrated that the Ti/W-based pMTJ nanodevices simultaneously exhibit an enhanced TMR ratio of 182%, a high-speed operation at nanosecond, and an ultra-low write energy $E_{sw}$ of less than 0.1 pJ/bit, thereby satisfying the trade-off between read-signal margin and power dissipation. We also revealed a projected room-temperature retention time exceeding 10 years, together with an endurance over $10^{12}$ cycles and a device yield surpassing 99.95%. These results convincingly show that the optimized Ti/W channel can collectively improve read margin, write efficiency and process tolerance. Our results constitute an important advance in resolving the longstanding challenges in the MRAM industry, which require a simultaneous realization of large TMR ratio, low write current, fast operation speed and CMOS-compatible manufacturability. Further optimization of the orbital-to-spin conversion efficiency in the low-cost light metals, together with the field-free switching could provide a viable and scalable pathway for promoting the deployment of OT-MRAM down to advanced technology nodes (e.g., sub-22 nm) with compelling area and energy efficiency, amongst many other advantages.

---

## Methods

**Sample preparation and characterization:** The top-pinned (TP) perpendicular magnetic tunnel junction (pMTJ) stacks were integrated on a 300 mm wafer platform with standard back-end-of-line (BEOL) processes. The Cu damascene bottom metal (BM) and W bottom via (BV) were constructed to form the electrical contact with the orbital torque (OT) and spin-orbit torque (SOT) channels. An optimized chemical mechanical planarization (CMP) process was performed to ensure a smooth BV interface. On top of the smooth BV surface, the pMTJ stacks consisted of channel layer (Ti/W, W, Ru/W, and Mo/W)/CoFeB (1.05 nm)/MgO (1.4 nm)/CoFeB (0.9 nm)/SAF layer were subsequently deposited by a multi-cathode magnetron sputtering PVD system under a base pressure better than $3 \times 10^{-8}$ Torr, followed by annealing at 350 °C in a high-vacuum furnace. The SAF layer is composed of W (0.4 nm)/[Co (0.7 nm)/Pt (1.0 nm)]$_2$/Co (0.6 nm)/Ru (0.45 nm)/[Co (0.6 nm)/Pt (0.25 nm)]$_7$ and capped by Ru (2.0 nm)/Ta (1.0 nm)/Ru (2.0 nm). The Ti ($x$ nm)/W ($y$ nm)/CoFeB ($z$ nm)/MgO (1.0 nm)/capping layer ($x$ = 4, 6, 8, 10 nm; $y$ = 0.5, 1.0, 1.5, 2.0 nm; $z$ = 1.05, 5 nm) were also prepared and examined under the same conditions.

To investigate the structural properties of the pMTJ stacks, we performed cross-sectional transmission electron microscopy (TEM) and energy-dispersive X-ray spectroscopy (EDS) studies. The cross-sectional TEM lamellae were prepared using a dual-beam focused-ion-beam (FIB) system (Helios G4 UX, Thermo Fisher Scientific). Structural and chemical analyses were carried out on a Talos F200X TEM (Thermo Fisher Scientific) that is equipped with an EDS detector, operating at 200 kV. Elemental mapping was performed to examine the composition and interfacial quality of the multilayer stacks. Spatially resolved electron energy-loss spectroscopic (EELS) mapping was performed on a 300-keV probe-corrected STEM FEI ThemisZ equipped with a Schottky field emitter and the Gatan 1066 energy filtering system.

**Device patterning:** The Ti ($x$ nm)/W ($y$ nm)/CoFeB ($z$ nm)/MgO (1.0 nm)/capping layer ($x$ = 4, 6, 8, 10 nm; $y$ = 0.5, 1.0, 1.5, 2.0 nm; $z$ = 1.05, 5 nm) were patterned into standard Hall bar devices with the dimensions of 20 × 60 $\mu m^2$ for spin-torque ferromagnetic resonance (ST-FMR) measurements and 8 × 60 $\mu m^2$ for current-induced magnetization switching measurements, which were done via a direct laser writing process using Ar-ion milling and photolithography method. The Ti (10 nm)/Pt (100 nm) contact electrodes were deposited by sputter deposition to ensure stable electrical contacts.

The hard mask, pMTJ stack and channel layers were patterned simultaneously by using an ion-beam etch (IBE) process with a critical dimension of ~ 60 nm (width) × 130 nm (length) on 300 mm wafers. Subsequently, the pMTJ stack and channel layers were encapsulated by dielectric (SiN) and followed by oxide filling. After another CMP process, the top electrode (TE) and a dual damascene Cu contact of top via (TV)/top metal (TM) together with the aluminium (Al) pads were formed to connect the top and bottom metal layers for the pMTJ devices. Note that the TMR ratio of the pMTJ nanodevices is slightly lower than that of the CIPT results of film, which could be attributed to mild degradation during the device fabrication process.

**Device measurements:** The anomalous Hall effect (AHE) and current-induced perpendicular magnetization switching were characterized using a custom-built electrical setup (Keithley 6221 current source and 2182 nanovoltmeter). AHE curves were recorded with a 1 mA probe current under out-of-plane magnetic fields ($H_z$), while magnetization switching was induced by current pulses and read out with 1 mA to capture the device switching behavior, in the presence of an in-plane magnetic field ($H_x$).

ST-FMR measurements were performed by applying 3-9 GHz RF signals (~23 dBm) along the long axis of the device. Angle-dependent measurements were conducted by changing the angle between the direction of field and the longitudinal axis of the device. The RF frequency with an optimal signal-to-noise ratio was selected for each device, and the torque efficiency ($\xi_{OT}$) was extracted from the angular dependence of the linewidth and amplitude of the ST-FMR signal.

The switching measurements of the pMTJ nanodevices were performed using a probe station with accesses to magnetic field and external electronics. A Keysight 81160A was used to generate square voltage pulses with a minimum pulse width of 2 ns. After each pulse being terminated, the resistance of the pMTJ nanodevices and the channels was probed using a Keysight B1500A. During the measurements, an in-plane magnetic field along the current direction ($H_x$) was applied. The endurance testing of the pMTJ nanodevices was performed using the same probe station and measurement setup as the magnetization switching of the pMTJ devices at room temperature. Sequences of positive and negative pulses generated by the Keysight 81160A, with a width of 5 ns, were applied at a voltage approximately 1.5 times larger than the critical switching voltage under the given in-plane $H_x$. The key parameters such as channel resistance ($R_{ch}$), parallel-state resistance ($R_P$) and antiparallel-state resistance ($R_{AP}$) are monitored by Keysight B1500A when the devices were subjected to a continuous series of pulses, specifically at $1\times10^n$, $3\times10^n$, $5\times10^n$, $7\times10^n$ cycles.

Data-retention extraction was carried out using a wafer prober at room temperature under various fixed perpendicular magnetic field ($H_z$). During the measurement, resistance variations of the pMTJ nanodevice array (in total 2500 nanodevices) were monitored to determine the switching probability, $P(H_z)$, with each measurement interval set to 10 s. The effective thermal stability factor at the corresponding magnetic field, $\Delta_H$, was extracted by fitting the time-dependent $P(H_z)$ curve through the equation of with $P(H_z) = 1 - \exp[-tf_0 \exp\left(\Delta_H - \frac{\sigma_\Delta{}^2}{2}\right)]$, where, attempt frequency $f_0$ equals 1GHz, $H_s$ is the offset field of the FL. Based on domain wall mediated reversal (DWMR) model[55,56] $\Delta_{\mathrm{H}} = \frac{1}{k_B T}\left\{\sigma_w R_{MTJ} t_{FL} + \frac{\sigma_w^2 t_{FL}}{2M_s(H_z - H_s)}\left[\frac{\pi}{2} - tan^{-1}\left(\frac{\sigma_w}{2M_s(H_z - H_s)R_{MTJ}}\right)\right] - 2M_s(H_z - H_s)t_{FL}R_{MTJ}^2 tan^{-1}\left(\frac{\sigma_w}{2M_s(H_z - H_s)R_{MTJ}}\right)\right\}$, the data retention $\Delta$ at zero magnetic field was obtained.

**Theoretical calculations:** The orbital Hall conductivity and spin Hall conductivity of *hcp* Ti and $\alpha$-W were calculated by first-principles calculations employing the Perdew-Burke-Ernzerhof generalized-gradient approximation. The electron-ion interactions were treated using the projector-augmented-wave method. A plane-wave kinetic-energy cutoff of 450 eV was employed for all three systems. The lattice parameters and atomic coordinates were fully relaxed until the residual Hellmann–Feynman force on each atom was below 1 meV Å$^{-1}$. The electronic self-consistency criterion was set to $1 \times 10^{-6}$ eV. K-point meshes of $15 \times 15 \times 10$ and $16 \times 16 \times 16$ were used for *hcp* Ti and $\alpha$-W, respectively. Wannier tight-binding Hamiltonians were constructed using Wannier90[63], with the s, p, and d atomic orbitals of Ti or W employed as the initial projections. The intrinsic spin Hall conductivity was evaluated within the Kubo-Berry-curvature formalism using maximally localized Wannier functions and WannierTools[64,65]. The intrinsic orbital Hall conductivity was calculated using an in-house extension of the Wannier tools workflow, following the first-principles Wannier-interpolation formalism for the orbital Hall effect[66]. The real-space Hamiltonian and the operator matrix elements required for the spin- and orbital-current responses were interpolated onto a dense $200 \times 200 \times 200$ k-point mesh for Brillouin-zone integration.

## Data Availability

The authors declare that the data supporting the findings of this study are available within the main text or Supplementary Information files. Additional data are available from corresponding authors upon request.

## Ethics declarations

Competing interests: The authors declare no competing interests.

# References

1. Dieny, B. et al. Opportunities and challenges for spintronics in the microelectronics industry. Nat. Electron. 3, 446–459 (2020).
2. Lanza, M. et al. Memristive technologies for data storage, computation, encryption, and radio-frequency communication. Science 376, eabj9979 (2022).
3. Lu, A. et al. High-speed emerging memories for AI hardware accelerators. Nat. Rev. Electr. Eng. 1, 24–34 (2024).
4. Khwa, W. S. et al. A mixed-precision memristor and SRAM compute-in-memory AI processor. Nature 639, 617–623 (2025).
5. Wan, W. et al. A compute-in-memory chip based on resistive random-access memory. Nature 608, 504–512 (2022).
6. Jung, S. et al. A crossbar array of magnetoresistive memory devices for in-memory computing. Nature 601, 211–216 (2022).
7. Park, S. O. et al. Phase-change memory via a phase-changeable self-confined nano-filament. Nature 628, 293–298 (2024).
8. Feng, G. et al. In-memory ferroelectric differentiator. Nat. Commun. 16, 3027 (2025).
9. Miron, I. M. et al. Perpendicular switching of a single ferromagnetic layer induced by in-plane current injection. Nature 476, 189–193 (2011).
10. Liu, L. et al. Spin–torque switching with the giant spin Hall effect of tantalum. Science 336, 555–558 (2012).
11. Nguyen, V. D. et al. Recent progress in spin-orbit torque magnetic random-access memory. npj Spintronics 2, 48 (2024).
12. Han, X. F. et al. Spin-orbit torques: Materials, physics, and devices. Appl. Phys. Lett. 118, 120502 (2021).
13. Song, C. et al. Spin-orbit torques: Materials, mechanisms, performances, and potential applications. Prog. Mater. Sci. **118,** 100761 (2021).
14. Fert, A. et al. Electrical control of magnetism by electric field and current-induced torques. Rev. Mod. Phys. 96, 015005 (2024).
15. Ji, G. et al. Recent progress on controlling spin-orbit torques by materials design. npj Spintronics 2, 56 (2024).
16. Demasius, K U. et al. Enhanced spin–orbit torques by oxygen incorporation in tungsten films. Nat. Commun. 7, 10644 (2016).
17. Sethu, K. K. V. et al. Optimization of Tungsten $\beta$-phase window for spin-orbit-torque magnetic random-access memory. Phys. Rev. Appl. 16, 064009 (2021).
18. Wu, H. et al. Magnetic memory driven by topological insulators. Nat. Commun. 12, 6251 (2021).
19. Grimaldi, E. et al. Single-shot dynamics of spin–orbit torque and spin transfer torque switching

in three-terminal magnetic tunnel junctions. Nat. Nanotechnol. 15, 111–117 (2020) .

20. Beek, S. V. et al. Scaling the SOT track-A path towards maximizing efficiency in SOT-MRAM. In International Electron Devices Meeting (IEDM) 1-4 (IEEE, 2023).
21. Xiang, Y. et al. SOT-MRAM bitcell scaling with BEOL read selectors: A DTCO study. IEEE Transactions on Electron Devices. 72, 12, 6665-6671(2025).
22. Cai, K. et al. First demonstration of field-free perpendicular SOT-MRAM for ultrafast and high-density embedded memories. In International Electron Devices Meeting (IEDM) 36.2.1-36.2.4 (IEEE, 2022).
23. Nguyen, V. D. et al. Achieving 1ppm write-error rate in SOT-MRAM with synthetic antiferromagnetic free layer. In IEEE International Electron Devices Meeting (IEDM) 1-4 (IEEE, 2024).
24. Li, K. -S. et al. First BEOL-compatible, 10 ns-fast, and durable 55 nm top-pSOT-MRAM with high TMR (>130%). In International Electron Devices Meeting (IEDM) 1-4 (IEEE, 2023).
25. Zeng, D. G. et al. High TMR over 156% in perpendicular SOT-MRAM realized with channel engineering. IEEE Electron Device Letters. 47, 2, 411-414 (2026).
26. Naik, V. B. et al. JEDEC-qualified highly reliable 22nm FD-SOI embedded MRAM for low-power industrial-grade, and extended performance towards automotive-Grade-1 applications. In IEEE International Electron Devices Meeting (IEDM) 11.3.1-11.3.4 (IEEE, 2020).
27. Yang, H. et al. Two-dimensional materials prospects for non-volatile spintronic memories. Nature 606, 663–673 (2022).
28. Gupta, R. et al. Harnessing orbital Hall effect in spin-orbit torque MRAM. Nat. Commun. 16, 130 (2025).
29. Xu, J. K. et al. Orbital-current-driven magnetization switching in a magnetic tunnel junction. Phys. Rev. Applied 25, 034023 (2026).
30. Yao, Y. X. et al. Giant orbital torque-driven picosecond switching in magnetic tunnel junctions, Sci. Bulletin. DOI: 10.1016/j.scib.2026.07.019 (2026).
31. Ding, S. L. et al. Generation, transmission, and conversion of orbital torque by an antiferromagnetic insulator. Nat Commun 16, 9239 (2025).
32. Go, D. et al. Intrinsic spin and orbital Hall effects from orbital texture. Phys. Rev. Lett. **121,** 086602 (2018).
33. Go, D. et al. Orbital torque: Torque generation by orbital current injection. Phys. Rev. Research 2, 013177 (2020).
34. Lee, D. et al. Orbital torque in magnetic bilayers. Nat. Commun. **12,** 6710 (2021).
35. Sala, G. & Gambardella, P. Giant orbital Hall effect and orbital-to-spin conversion in 3d, 5d, and 4f metallic heterostructures. *Phys. Rev. Res.* **4,** 033037 (2022).
36. Wang, P. et al. Inverse orbital Hall effect and orbitronic terahertz emission observed in the materials with weak spin-orbit coupling. npj Quantum Mater. **8**, 28 (2023).
37. Choi, Y. G. et al. Observation of the orbital Hall effect in a light metal Ti. Nature 619, 52–56

(2023).

38. Hayashi, H. et al. Observation of orbital pumping. Nat. Electron. 7, 646-652 (2024).
39. Schmitt, C. et al. Orbital magnetoresistance in the antiferromagnet CoO driven by dynamic orbital angular momentum. Science 393,76-79(2026).
40. Xie, H. et al. Efficient noncollinear antiferromagnetic state switching induced by the orbital Hall effect in chromium. *Nano Lett.* **23,** 10274-10281(2023).
41. Fukunaga, R. et al. Orbital torque originating from orbital Hall effect in Zr. Phys. Rev. Research 5, 023054 (2023).
42. Yang, Y. et al. Orbital torque switching in perpendicularly magnetized materials. Nat. Commun. 15, 8645 (2024).
43. Zheng, Z. et al. Effective electrical manipulation of a topological antiferromagnet by orbital torques. *Nat. Commun.* **15,** 745 (2024).
44. Zhang, D. L. et al. Orbital torque switching of room temperature two-dimensional van der waals ferromagnet $Fe_3GaTe_2$. Nat. Commun. **16**, 7047 (2025).
45. Shin, S. et al. Enhanced magnetization switching efficiency via orbital-current-induced torque in Ti/Ta (Pt)/CoFeB/MgO structures. Adv. Funct. Mater. 2425932 (2025).
46. Fukami, S., Lee, K. J. & Kläui, M. Challenges and opportunities in orbitronics. Nat. Phys. (2025).
47. Buter, W. H. et al. Spin-dependent tunneling conductance of Fe|MgO|Fe sandwiches. Phys. Rev. B 63, 054416 (2001).
48. Niessen, A. K., & De Boer, F. R. J. Less-Common Met. 82, 75 (1981).
49. Guan, T. et al. Evidences of subnanometre orbital diffusion length in heavy metals using terahertz emission spectroscopy. Nat. Nanotechnol. 21, 538–545 (2026).
50. Zhao, L. et al. Manufacturing-friendly SOT-MTJ device with high reliability and switching efficiency. IEEE Electron Device Letters. 46, 8, 1345-1348 (2025).
51. Yuasa S. et al. Giant room-temperature magnetoresistance in single-crystal Fe/MgO/Fe magnetic tunnel junctions. Nat. Mater. 3, 868-871 (2004).
52. Liu, E. L. et al. A novel channel-less SOT-MRAM with 115% TMR, 2 ns switching, and high bit yield (>99.9%). In IEEE International Electron Devices Meeting (IEDM) 1-4 (IEEE, 2024).
53. Wu, S. Y. et al. A 7nm CMOS platform technology featuring 4th generation FinFET transistors with a 0.027 $um^2$ high density 6-T SRAM cell for mobile SoC applications. In IEEE International Electron Devices Meeting (IEDM) 1-4 (IEEE, 2016).
54. Dieny, B., & Chshiev, M. Perpendicular magnetic anisotropy at transition metal/oxide interfaces and applications. Rev. Mod. Phys. 89, 025008 (2017).
55. Mihajlović, G. et al. Thermal stability for domain wall mediated magnetization reversal in perpendicular STT MRAM cells with W insertion layers. Appl. Phys. Lett. 117, 242404 (2020).
56. Chaves-O'Flynn, G. D. et al. Thermal stability of magnetic states in circular thin-film nanomagnets with large perpendicular magnetic anisotropy. Phys. Rev. Applied 4, 024010

(2015).

57. Sato, N. et al. CMOS compatible process integration of SOT-MRAM with heavy-metal bi-layer bottom electrode and 10ns field-free SOT switching with STT assist. In IEEE Symposium on VLSI Technology 1-2 (IEEE, 2020).
58. Song, M. Y. et al. High speed (1ns) and low voltage (1.5V) demonstration of 8Kb SOT-MRAM array. In IEEE Symposium on VLSI Technology and Circuits 377-378 (IEEE, 2022).
59. Jiang, C. P. et al. Demonstration of 128 Kb SOT-MRAM chip with 5 ns Write and 15 ns read speed, high endurance over $10^{10}$ and low ECC-on bit error rate. In IEEE International Electron Devices Meeting (IEDM) 1-4 (IEEE, 2024).
60. Nguyen, T. V. A. et al. Low write power and field-free sub-ns write speed SOT-MRAM cell with design technology of canted SOT structure and magnetic anisotropy for NVM. In IEEE International Memory Workshop (IMW) 1-4 (IEEE, 2025).
61. Huang, Y. L. et al. A 64-kilobit spin-orbit torque magnetic random-access memory based on back-end-of-line-compatible $\beta$-tungsten. Nat. Electron. 8, 794–802 (2025).
62. Iwai, H. Roadmap for 22 nm and beyond. Microelectronic Engineering. 86, 1520-1528 (2009).
63. Mostofi, A. A. et al. Wannier90: A tool for obtaining maximally-localised Wannier functions. Comput. Phys. Commun. 178, 685–699 (2008).
64. Wu, Q. S. et al. WannierTools: An open-source software package for novel topological materials. Comput. Phys. Commun. 224, 405–416 (2018).
65. Qiao, J. et al. Calculation of intrinsic spin Hall conductivity by Wannier interpolation. Phys. Rev. B 98, 214402 (2018).
66. Go, D. et al. First-principles calculation of orbital Hall effect by Wannier interpolation: Role of orbital dependence of the anomalous position. Phys. Rev. B 109, 174435 (2024).

**Acknowledgements**

We acknowledge the support from team members of Hikstor MRAM pilot line. D.L.Z. and Y.J. gratefully acknowledges the research funding provided by the National Key R&D Program of China (2022YFA1204003) and the National Natural Science Foundation of China (Grant Nos. 52271240, U23A20551, U24A6002). Work carried out at Tsinghua was supported by the distinguished Young Scholar program of National Natural Science Foundation of China (NSFC Grant No. 12225409), Basic Science Center Project (NSFC Grant No. 52388201), the National Key R&D Program of China (Grant No. 2022YFA1405100), the NSFC general program (Grant Nos. 52271181, 12421004, 12404138).

**Author contributions:** D.G.Z., Y.G., and J.Y.D. contributed equally to this work. D.G.Z. and S.K.H. initialized and conceived this work. S.K.H. coordinated and supervised the project. D.G.Z., D.L.Z., and Y.G. conceived the experiments. J.Y.D., Y.H.A., D.L.Z., and Y.J. patterned the Hall bar devices, carried out the ST-FMR & second Harmonic Hall measurements, performed the crystalline characterizations, and calculated the spin Hall conductivity and orbital Hall conductivity. S.S.W., L.Z., and Z.H.J. developed the SARR process integration flow and pMTJ etch process on 300 mm wafers. H.A.Z. performed device retention test. L.Z., and X.J.X. performed MRAM circuit-level bitcell simulations. X.H., J.T.K., Y.L.G., M.Y.C., E.L.L., J.L.G., Q.J.G., Y.H.S., Z.J.Z., W.M.H., X.L.Y., F.T.M., Y.H.W., H.X.Y., and W.J.J coordinated resources and analyzed the experimental data. D.G.Z., D.L.Z., W.J.J., and S.K.H wrote the manuscript. All the authors discussed the results and commented on the manuscript.